\documentclass[aps,twocolumn,superscriptaddress]{revtex4}
\usepackage{amsmath}
\usepackage{amsfonts}
\usepackage{mathrsfs}
\usepackage{graphicx}
\usepackage{times}
\usepackage{color}

\def\be{\begin{equation}}
\def\ee{\end{equation}}
\def\bea{\begin{eqnarray}}
\def\eea{\end{eqnarray}}

\begin{document}

\title{Chiral orbital magnetism of $p$-orbital bosons in optical lattices}

\author{Bo Liu}
\email{liubophy@gmail.com}
\affiliation{Department of Applied Physics, School of Science,
Xi'an  Jiaotong University, Xi'an  710049, Shaanxi, China}
\affiliation{Shaanxi Province Key Laboratory of Quantum Information and Quantum Optoelectronic Devices, Xi'an Jiaotong University, Xi'an 710049, Shaanxi, China}
\author{Peng Zhang}
\affiliation{Department of Applied Physics, School of Science,
Xi'an  Jiaotong University, Xi'an  710049, Shaanxi, China}
\author{Hong Gao}
\affiliation{Department of Applied Physics, School of Science,
Xi'an  Jiaotong University, Xi'an  710049, Shaanxi, China}
\affiliation{Shaanxi Province Key Laboratory of Quantum Information and Quantum Optoelectronic Devices, Xi'an Jiaotong University, Xi'an 710049, Shaanxi, China}
\author{Fuli Li}
\affiliation{Department of Applied Physics, School of Science,
Xi'an  Jiaotong University, Xi'an  710049, Shaanxi, China}
\affiliation{Shaanxi Province Key Laboratory of Quantum Information and Quantum Optoelectronic Devices, Xi'an Jiaotong University, Xi'an 710049, Shaanxi, China}

\begin{abstract}

Chiral magnetism is a fascinating quantum phenomena that has been found in low-dimensional magnetic materials.
It is not only interesting for understanding the concept of chirality, but also important for potential applications
in spintronics. Past studies show that chiral magnets require both lack of the inversion symmetry and spin-orbit coupling to induce the Dzyaloshinskii-Moriya (DM) interaction. Here we report that the combination of inversion symmetry breaking and quantum degeneracy of orbital degrees of freedom will provide a new paradigm to achieve the chiral orbital magnetism.
By means of the density matrix renormalization group (DMRG) calculation, we demonstrate that the chiral orbital magnetism
can be found when considering bosonic atoms loaded in the $p$-band of an optical lattice in the Mott regime. The high tunability of our scheme is also illustrated through simply manipulating the inversion symmetry of the system for the
cold atom experimental conditions.

\end{abstract}

\maketitle

Chirality plays an important role in physics and influences many of the physical properties in a profound way.
In condensed matter physics, the search for non-trivial chiral spin textures is currently of great interest~\cite{2013_Nagaosa_Review},
such as in non-centrosymmetric magnetic metals~\cite{2006_Uchida_science,2006_Roszler_nature} and heterostructure thin-film materials~\cite{2015_Mihai_Review,2016_Fukami_NatMat}. Understanding chiral
magnetic order and its dynamics driven by magnetic fields is important not only for exploring the variety of
fascinating phenomena resulting from the topologically protected magnetic structures, such as
Hall effects and other transport features~\cite{2009_Binz_PhysRevLett,2006_Binz_PhysRevLett,2006_Binz_PhysRevB}, but also to help unlock the potential of novel spintronic technologies~\cite{2008_Parkin_science,2005_Allwood_science,2016_Woo_NatMat}.
However, how to manipulate chiral spin textures is still a significant challenge in solid state materials~\cite{2008_Parkin_science}.
Besides the continuously growing effort to study the magnetic chirality in solids, there have been great interests of exploring the chiral magnetic textures using ultracold atoms in both experimental and theoretical studies~\cite{2012_shizhong_PhysRevLett,2014_Shizhong_PhysRevA,2014_Zicai_PhysRevA}, motivated
by the recent experimental advances to create tunable spin-orbit coupling or nonabelian gauge fields in general~\cite{2013_Galitski_nature,2011_RevModPhys,2011_Lin_nature,2016_Wu_science,2012_Cheuk_PhysRevLett,2012_Wang_PhysRevLett,
2014_Jotzu_Nature,2014_Bloch_science,2013_Aidelsburger_PhysRevLett,2013_Miyake_PhysRevLett,2013_cheng_Natphys}.
Such highly controllable atomic systems will not only provide a versatile tool for simulating the magnetic chirality in electronic systems, but also supply new probabilities to manipulate the chiral magnetic textures with no counterpart in solids. However, the past studies in the cold atom based system require spin-orbit coupled strongly interacting atoms to induce the Dzyaloshinskii-Moriya (DM) interaction~\cite{2012_shizhong_PhysRevLett,2014_Shizhong_PhysRevA,2014_Zicai_PhysRevA}, for which future experimental breakthroughs are desired, in particular like to suppress heating and meanwhile achieve the strongly interacting regime~\cite{2013_Galitski_nature}.

Here we report the discovery of a new mechanism to achieve chiral magnetism constructed via orbital degrees of freedom. We shall introduce this with a specific model of cold bosonic atoms in an optical lattice, to be illustrated below. The key idea here is to introduce the non-trivial hybridization between degenerate orbital degrees of freedom via manipulating the inversion symmetry of the system. Surprisingly, such interplay between the inversion symmetry breaking and orbital degeneracy of interacting bosons in optical lattices provides a new scheme towards discovering chiral magnetism. The present mechanism departs from the conventional wisdom to realize chiral magmatic textures that relies on both lack of the inversion symmetry and spin-orbit coupling. Moreover, typically the degenerate orbitals could emerge in presence of point group symmetries, where the symmetry for orbitals is much lower than that for spins~\cite{2000_Tokura_science}. Therefore, this mechanism could provide an easier way to manipulate the chiral magnetism constructed through orbitals compared to build on from the spins in solid magnetic materials and also would shed light on the potential applications analogous to those in spintronic technologies via manipulation of the chiral orbital magnetism. This idea is motivated by the recent experimental progress in manipulating higher orbital bands in optical lattices~\cite{2016_Hemmerich_Review,2016_Vincent_Review}, like from the early experimental attempt to the breakthrough observation of long-lived $p$-band bosonic atoms in a checkerboard lattice pointing to an exotic $p_x \pm ip_y$ orbital Bose-Einstein condensate~\cite{2011_Hemmerich_NatPhys,2015_Hemmerich_PhysRevLett}. It provides unprecedented opportunities to investigate quantum many-body phases with orbital degrees of freedom~\cite{2014_Bo_arxiv,2016_Bo_PhysRevA,2016_Bo_PhysRevAI,2013_xiaopeng_natcomm,2010_Zixu_PhysRevA,
2011_Zicai_PRA,2012_Zicai_PhysRevB}. Ferro- and Antiferro- orbital order have been discussed in the pervious theoretical studies for single or multiple components fermions ~\cite{2008_congjun_PhysRevLett,2008_erhai_PhysRevLett,2014_Liyi_PhysRevLett,2015_erhai_PhysRevLett} or bosons~\cite{2013_Pinheiro_PhysRevLett,2012_xiaopeng_PhysRevLett} on the $p$-band. As we shall show with the model below, the inversion symmetry controlled non-trivial hybridization between degenerate orbitals can lead to other unexpected results.

\textit{Effective model $\raisebox{0.01mm}{---}$} Let us consider a gas of interacting bosonic atoms loaded in a 1D optical lattice which can be realized from a strongly anisotropic 2D square optical lattice. In particular, we consider the lattice potential $V_{\rm OL}({\mathbf r})=-V_x \cos^2(k_{Lx}x)-V_y \cos^2(k_{Ly}y)$ with lattice strengths $V_{y}>>V_{x}$ to achieve the 1D limit, where $k_{Lx}$ and $k_{Ly}$ are the wavevectors of the laser fields and the corresponding lattice constants are defined as $a_x=\pi/k_{Lx}$ and $a_y=\pi/k_{Ly}$ in the $x$ and $y$ directions respectively. In the deep lattice limit, the lattice potential at each site can be approximated by a harmonic oscillator. Under this approximation, to keep local rotation symmetry ~\cite{2012_xiaopeng_PhysRevLett} of each site in the $xy$-plane requires that the lattice potential satisfies the conditions $V_xk^2_{Lx}=V_yk^2_{Ly}$, which guarantees the twofold degeneracy of $p$-orbitals at each lattice site. These two degenerate $p$-orbitals (i.e., $p_x$ and $p_y$ orbitals) are well separated in the energy approximately by the harmonic oscillator frequency $\hbar\omega=\sqrt{4V_{x}E_{Rx}}=\sqrt{4V_{y}E_{Ry}}$ ($E_{Rx}={\hbar^2 k^2_{Lx}}/{2m}$ and $E_{Ry}={\hbar^2
k^2_{Ly}}/{2m}$ are the recoil energy) from other orbitals. The new ingredient of our model is the presence of a gradient magnetic field~\cite{2013_Aidelsburger_PhysRevLett,2013_Miyake_PhysRevLett} along the $y$-direction and the single-particle physics can be described through the following Hamiltonian $H_0=-\frac{\hbar^2}{2m}\nabla^2+V_{\rm OL}(\mathbf {r})-\mathbf{F}\cdot \mathbf {r}$, where $\mathbf{F}=-J\nabla_y B$ is the force applied to the atom with spin magnetic moment $J$ which is only along the $y$-direction. Here we want to emphasize the crucial role of the external gradient magnetic field in our proposal. First, it breaks the inversion symmetry along the $y$-direction and thus induces the non-trivial hybridization between the degenerate orbitals, i.e., $p_x$ and $p_y$ orbitals. Later on we shall see that this band mixing paired with the interaction between bosons will lead the orbital Dzyaloshinskii-Moriya interaction to further form the chiral orbital magnetism. Secondly, tunneling in the $y$-direction is further suppressed by a linear tilt of the energy per lattice site to make the system dynamically as 1D. A system of interacting spinless bosons loaded in these $p_x$ and $p_y$ orbitals can be described by the following
multi-orbital Bose Hubbard model in the tight binding regime
\begin{eqnarray}
\hat{H}&=&\sum_{\mathbf{i}}t_{x}\hat{b}_{p_{x}}^{\dag }(\mathbf{x_i})\hat{b}%
_{p_{x}}(\mathbf{x_i}+\vec{e}_{x})-\sum_{\mathbf{i}}t_{y}\hat{b}_{p_{y}}^{\dag }(%
\mathbf{x_i})\hat{b}_{p_{y}}(\mathbf{x_i}+\vec{e}_{x})\notag \\
&-&t\sum_{\mathbf{i}}[\hat{b}%
_{p_x}^{\dag }(\mathbf{x_i})\hat{b}_{p_{y}}(\mathbf{x_i}+\vec{e}_{x})
-\hat{b}_{p_{y}}^{\dag }(\mathbf{x_i})\hat{b}_{p_{x}}(\mathbf{x_i}+\vec{e}
_{x})] \notag \\
&+&h.c.+\sum_{{\substack{\mathbf{i}, \alpha=p_{x},p_{y}}}}[ \frac{U_{\alpha \alpha }}{2}
\hat{n}_{\alpha }(\mathbf{x_i})(\hat{n}_{\alpha }(\mathbf{x_i})-1)] \notag \\
&+&\sum_{{\substack{\mathbf{i}, \alpha=p_{x},p_{y} \\  \alpha^{\prime }=p_{x},p_{y},
\alpha \neq \alpha ^{\prime }}}}[U_{\alpha \alpha
^{\prime }}\hat{n}_{\alpha }(\mathbf{x_i})\hat{n}_{\alpha ^{\prime }}(\mathbf{x_i}) \notag \\
&+& \frac{U_{\alpha \alpha ^{\prime }}}{2}\hat{b}_{\alpha }^{\dagger }(\mathbf{x_i})\hat{b}
_{\alpha }^{\dagger }(\mathbf{x_i})\hat{b}_{\alpha ^{\prime }}(\mathbf{x_i})\hat{b}_{\alpha
^{\prime }}(\mathbf{x_i})]
\label{HamHubbard}
\end{eqnarray}
where $\hat{b}_{p_{x}}(\mathbf{x_i})$ ($\hat{b}_{p_{y}}(\mathbf{x_i})$)
is the annihilation operator for the bosonic particle
in the $p_{x}$ ($p_{y}$) orbital at lattice site $\mathbf{x_i}$. The onsite particle number
operator is defined as $\hat{n}_{\alpha=p_{x},p_{y}}(\mathbf{x_i})=\hat{b}_{\alpha }^{\dag }(\mathbf{x_i})\hat{b}_{\alpha
}(\mathbf{x_i})$ and $t_{x}$ is the longitudinal hopping of $p_x$ bosons, and $t_y$
is the transverse hopping of $p_y$ bosons. The relative sign of the hopping amplitude is
fixed by the parity of $p_x$ and $p_y$ orbitals. The interacting strength is given by
$U_{\alpha \alpha ^{\prime }}=U_{0}\int d\mathbf{x}\,|w^{\alpha }(\mathbf{x-x}_{i})|^{2}|w^{\alpha ^{\prime }}(\mathbf{x-x}_{i})|^{2}$, where $U_{0}>0$ is the onsite interaction strength determined by the scattering
length and $w^{\alpha(\alpha ^{\prime })}(\mathbf{x-x}_{i})$ is the Wannier function of orbital $%
\alpha(\alpha^{\prime })$ at lattice site $\mathbf{x}_{i}$. There are two key ingredients in our model: $1)$ the hybridization between $p_x$ and $p_y$ orbitals arises from the asymmetric shape of the $p_y$ orbital
wavefunction induced by the inversion symmetry breaking in the $y$-direction via adding a $y$-direction gradient magnetic field; $2)$ a special type interaction between spinless bosons (the last term in Eq.~\eqref{HamHubbard}) describing the flipping of a pair of bosonic atoms from the state $\alpha$ to the state $\alpha ^{\prime }$. Note that this term is absent in the case of the bosons in the lowest band of an optical lattice. As we shall show below, such
combination effect will lead a new mechanism to achieve the chiral orbital magnetism (see details in the Supplementary Material).

\begin{figure*}
  \begin{center}
  \includegraphics[scale=0.175]{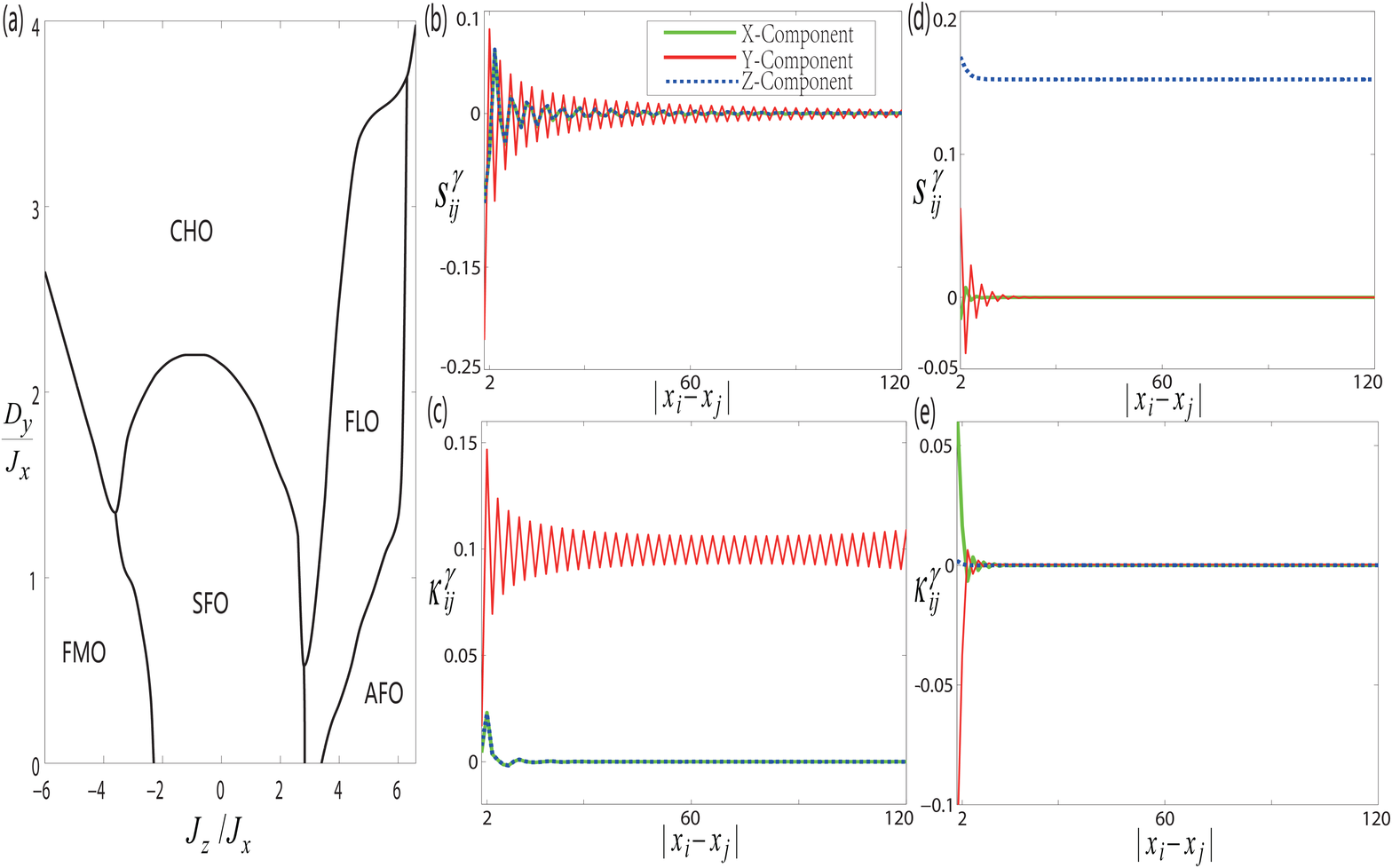}
  \end{center}
  \caption{(a) Phase diagram of the effective spin model in Eq.~\eqref{HMottSpinFinal}
   as a function of the orbital exchange interactions $D_y/J_x$ and $J_z/J_x$.
   For certain $J_z/J_x$ there is a threshold of the orbital DM interaction
   $D_y/J_x$, beyond that the chiral orbital (CHO) phase appears. Other parameters are
   $J_y/J_x=2.8$ and $h/J_x=0.2$. (b),(c) and (d),(e) show the spin-spin correlation $\mathcal{S}^{\gamma}_{ij}$
   and chiral correlation $\mathcal{K}^{\gamma}_{ij}$ for the chiral orbital (CHO) ($J_z/J_x=1$, $D_y/J_x=2.2$)
   and ferromagnetic orbital (FMO) ($J_z/J_x=-3.5$, $D_y/J_x=0.85$) phases, respectively. Other parameters are the
   same as in (a) and the system size is $L=150$ with the open boundary condition.}\label{fig:phase}
\end{figure*}

\textit{Orbital magnetism in the Mott insulator regime $\raisebox{0.01mm}{---}$} In this work, we will focus on the case of the Mott insulator phase with unit filling when considering the strongly repulsive limit, i.e.,$U_{\alpha \alpha ^{\prime}} \gg t_{x},t_{y},t$, to demonstrate that the chiral orbital magnetism can be produced from the orbital exchange interactions via the model (Eq.~\eqref{HamHubbard}) as proposed above. In our case, the orbital exchange interactions come from the virtual hopping processes, which can be described in terms of an effective Hamiltonian obtained from the perturbative expansion of the tunneling processes up to second order. Here, by introducing the projection operator~\cite{1998_book,2005_book} $\hat{P}$ to describe the sub-Hilbert space of the Mott space with singly occupied state and the projection onto the perpendicular subspace via the operator $\hat{Q}=1-\hat{P}$ correspondingly, the effective Hamiltonian which describes the one particle Mott phase of $p$-orbital bosons can be expressed as $\hat{H}_{eff}=-\hat{P}\hat{H}_{t}\hat{Q}\frac{1}{\hat{Q}\hat{H}_{U}\hat{Q}-E}
\hat{Q}\hat{H}_{t}\hat{P}$ where $\hat{H}_{U}$ and $\hat{H}_{t}$ are the interaction and hopping part
of the Hamiltonian in Eq.~\eqref{HamHubbard}. The details of derivation are given in the Supplementary Material.
Through constructing the pseudo-spin operators from the orbital degrees of freedom via $\hat{S}_i^{z}={\frac{1}{2}[\hat{b}_{p_x}^{\dagger }(\mathbf{x_i})\hat{b}_{p_x}(\mathbf{x_i})-\hat{b}_{p_y}^{\dagger }(\mathbf{x_i})\hat{b}_{p_y}(\mathbf{x_i})]}$, $\hat{S}_i^{+}=\hat{b}_{p_x}^{\dagger}(\mathbf{x_i})\hat{b}_{p_y}(\mathbf{x_i})$, $\hat{S}_i^{-}=\hat{b}_{p_y}^{\dagger }(\mathbf{x_i})\hat{b}_{p_x}(\mathbf{x_i})$
and together with the constraint of unit occupation of the lattice sites $\hat{n}_{p_x}(\mathbf{x_i})+\hat{n}_{p_y}(\mathbf{x_i})=1$,
the effective Hamiltonian can be mapped onto a pseudo-spin-${1\over 2}$ XYZ model with
orbital Dzyaloshinskii-Moriya (DM) interaction in an external field:
\begin{eqnarray}
\hat{H}_{eff}&=&\sum_{\langle ij \rangle} J_{x}\hat{S}_{i}^{x}\hat{S}_{j}^{x}+\sum_{\langle ij \rangle}J_{y}\hat{S}_{i}^{y}\hat{S}_{j}^{y}+\sum_{\langle ij\rangle}J_{z} \hat{S}_{i}^{z}\hat{S}_{j}^{z}  \notag \\
&+&\sum_{\langle ij\rangle }
\mathbf{D \cdot (\hat{\mathbf{S}}_{i} \times \hat{\mathbf{S}}_{j})}+h\sum_{i}\hat{S}_{i}^{z}  \label{HMottSpinFinal}
\end{eqnarray}
where $J_x=\frac{2(t_{x}t_{y}+t^2)}{U_{p_xp_y}}-\frac{8(t^{2}+t_{x}t_{y})U_{p_xp_y}}{U^{2}}$,
$J_y=\frac{2(t_{x}t_{y}-t^2)}{U_{p_xp_y}}+\frac{8(t_{x}t_{y}-t^{2})U_{p_xp_y}}{U^{2}}$,
$J_z =\frac{-4(t_{x}^{2}U_{p_yp_y}+t_{y}^{2}U_{p_xp_x})+4t^{2}(U_{p_xp_x}+U_{p_yp_y})}{U^{2}}-
\frac{2t^{2}-(t_{x}^{2}+t_{y}^{2})}{U_{p_xp_y}}$ with $U^{2}=U_{p_xp_x}U_{p_yp_y}-U_{p_xp_y}^{2}$
and the strength of an effective magnetic field $h=\frac{4t_{y}^{2}U_{p_xp_x}-4t_{x}^{2}U_{p_yp_y}}{U^{2}}$.
The fourth term in Eq.~\eqref{HMottSpinFinal} is the defined orbital
Dzyaloshinskii-Moriya (DM) interaction term here, with a DM vector $\mathbf{D}=(0,D_{y},0)$ and  $D_{y}=\frac{4tt_{x}(U_{p_yp_y}-U_{p_xp_y})+4tt_{y}(U_{p_xp_y}-U_{p_xp_x})}{U^{2}}$. It is not only strongly
reminiscent of its counterpart in strongly correlated electronic materials, such as in the cuprate superconductor
YBa$_2$Cu$_3$O$_6$ or in low-dimensional magnetic materials, but more importantly, constructing DM interaction
from orbital degrees of freedom (orbital DM interaction) would show some unique properties. For example,
due to the interplay of non-trivial hybridization between the degenerate orbitals (the term proportional to $t$ in Eq.~\eqref{HamHubbard}) induced by the presence of external magnetic gradient and the unequal inter-orbital
and intra-orbital interacting strength, i.e., $U_{p_xp_x(p_yp_y)}\neq U_{p_xp_y}$ resulting from the anisotropic shape of  $p$-orbitals, the orbital exchange interactions in Eq.~\eqref{HMottSpinFinal} are strongly anisotropic.
Unlike the $1D$ isotropic Heisenberg model, where the additional DM interaction
can be gauged away by performing a spin rotation~\cite{1992_Shekhtman_PhysRevLett}, the orbital DM interaction proposed here in Eq.~\eqref{HMottSpinFinal}
can no longer be gauged away. It will play an important role in determining the magnetic properties of the system as
illustrated below.

We investigate the phase diagram numerically by computing the
ground state of Hamiltonian in Eq.~\eqref{HMottSpinFinal} through
the density-matrix renormalization group
(DMRG) method~\cite{1992_white_PhysRevLett,2005_schollw_RevModPhys}, which has been very successfully used to explore
various properties of one-dimensional correlated systems.
In our calculations, we impose an open boundary condition
and use a finite-size DMRG algorithm. A finite-size scaling
analysis of different system sizes and extrapolation to the
thermodynamic limit is taken to identify the phase transition point.
In the application to the model Hamiltonian in Eq.~\eqref{HMottSpinFinal},
it is necessary to truncate the Hilbert space for each site.
For the present computation, the total sizes are up to $L =250$ lattice
sites with open boundary conditions and we keep up to $1000$
states in the matrix product state representation. We also utilized the
finite-size algorithm with $4$ to $10$ sweeps to reach the convergence
and the truncation error of the reduced density matrix is typically up
to $10^{-8}$ and the energies are converged up to seventh digit for the
ground-state energy per site.

The phase diagram of Hamiltonian in Eq.~\eqref{HMottSpinFinal} is presented
in Fig.~\ref{fig:phase}(a). There are five different magnetic phases obtained within
the DMRG calculations as shown in the phase diagram, which consists of a gapped
ferromagnetic orbital (FMO) phase, a gapless chiral orbital (CHO) phase, a gapped
antiferromagnetic orbital (AFO) phase, a gapped spin-flop orbital (SFO) phase
(corresponding to a gapped phase with N\`{e}el order in the $y$-component of
the pseudo-spin defended in Eq.~\eqref{HMottSpinFinal}), and a floating~\cite{2011_SelaPhysRevB,1982_Bak_report}
orbital (FLO) phase (a gapless phase without long range order and for which the correlations
decay algebraically). Here we will focus on the important
role of the orbital DM interaction in determining the magnetic properties of the system.
As shown in Fig.~\ref{fig:phase}(a), there is a threshold strength of the orbital DM
interaction to separate the magnetic phase with non-trivial chirality from other phases
in the phase diagram. Below that threshold, for large positive values of ${J_z}/{J_x}$, the ground state of the system
is AFO phase in the $z$-component of the pseudo-spin, corresponding to a staggered orbital
structure (within the two nearest-neighbor sites, one is occupied by the $p_x$ orbital and
the other is the $p_y$ orbital). When increasing the strength of the orbital DM interaction,
a first order phase transition occurs between AFO phase and CHO phase. For small values of
${J_z}/{J_x}$ (considering the case with ${J_y>J_x}$ here), below the
threshold the ground state of the system is featured with the N\`{e}el order
in the $y$-component of the pseudo-spin and the system is in the SFO phase.
Above the threshold strength, there is a first order phase transition between SFO phase
and CHO phase. Due to the anisotropic orbital exchange interaction in Eq.~\eqref{HMottSpinFinal},
below the threshold, there is a FLO phase between AFO phase and SFO phase, where
the correlations are characterized by the power-law decay. The transition from the
AFO phase to the FLO phase belongs to the commensurate-incommensurate type and
the Berezinsky-Kosterlitz-Thouless transition occurs between FLO and SFO phases.
Above the threshold, a continuous phase transition emerges between FLO and CHO phases.
When considering large negative values of ${J_z}/{J_x}$, below the threshold the
ground state of the system is in a FMO phase in the $z$-component of the pseudo-spin, corresponding
to a polarized orbital structure. And the transition between FMO phase and SFO phase is of Ising type.
Above the threshold strength, there is a first order phase transition between FMO phase and CHO phase.

\begin{figure}[t]
\begin{center}
\includegraphics[scale=0.13]{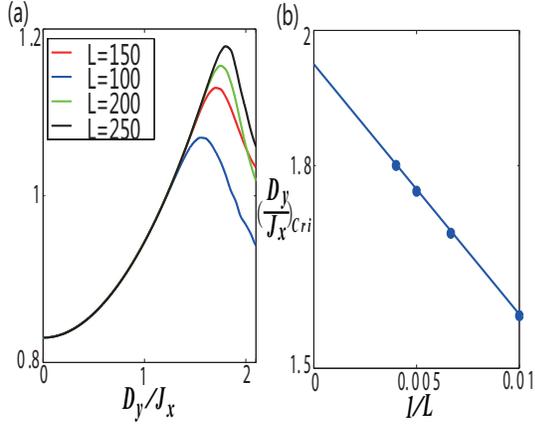}
\end{center}
\caption{(a)Entanglement entropy $S_{\rm E}$ as a function of $D_y/J_x$ with different system size $L$, calculated with open boundary condition when $J_y/J_x=2.8$, $J_z/J_x=1$ and $h/J_x=0.2$. The peak position of $D_y/J_x$ determines the phase transition point $(D_y/J_x)_{\rm Cri}$. (b) shows the finite size scaling of $(D_y/J_x)_{\rm Cri}$.} \label{fig:entropy}
\end{figure}

\begin{figure}[t]
\begin{center}
\includegraphics[scale=0.13]{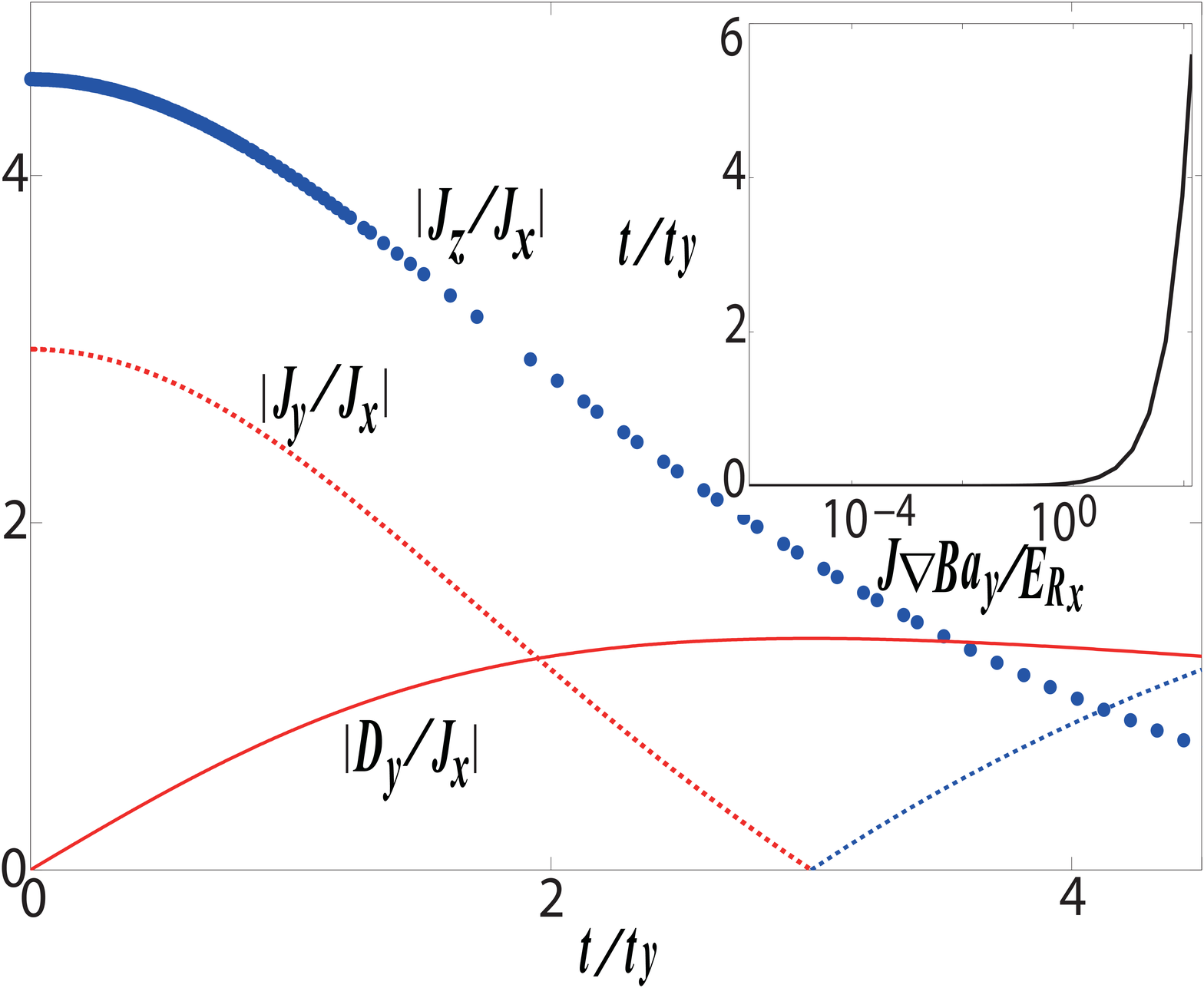}
\end{center}
\caption{The orbital exchange interactions as a function of the orbital hybridization $t/t_y$ when $V_{x}/E_{Rx}=6$ and
  $V_{y}/E_{Ry}=24$. Inset shows the orbital hybridization $t/t_y$ as a function of the gradient magnetic field. Here
  the red and blue color label the sign (positive and negative) of orbital exchange interaction, respectively.} \label{fig:coupling}
\end{figure}

Now let us discuss the difference among these five magnetic phases in terms of spin-spin correlation function
and chiral correlation function. Here, the spin-spin correlation function is defined as  $\mathcal{S}^{\gamma}_{ij} = \langle\psi_0| S^{\gamma}_{i} S^{\gamma}_{j}|\psi_0\rangle$, whereas the chiral correlation
function is defined as $\mathcal{K}^\gamma_{ij}\equiv\langle\psi_0| \mathcal{K}_i^{\gamma}\mathcal{K}_j^{\gamma}|\psi_0\rangle$ with $\gamma=x,y,z$ and $\mathcal{K}_i^{\gamma}=\varepsilon^{\gamma \mu \nu}(S_i^{\mu}S_{i+1}^{\nu}-S_i^{\nu}S_{i+1}^{\mu})$, where $|\psi_0\rangle$ is the ground state of the system. The AFO, SFO and FMO phases can be distinguished
by analyzing the asymptotic behavior of spin-spin correlation function. These phases are characterized by a non-zero
asymptotic value of the spin-spin correlation $\mathcal{S}^{\nu}\equiv\lim_{|i-j|\to\infty}\mathcal{S}^{\nu}_{ij}$.
For example, as shown in Fig. ~\ref{fig:phase}(d) and (e), the FMO phase shows a non-zero spin-spin correlation in the $z$-component of the pseudo-spin when considering large $|i-j|$ limit. However, it is quite different in
the CHO phase. As shown in Fig. ~\ref{fig:phase}(b) and (c), the spin-spin correlation function oscillates and their envelope function decays algebraically, but the chiral correlation demonstrates a non-zero asymptotic value, i.e., the CHO phase with finite $y$-component of chiral correlation $\mathcal{K}^{y}\equiv \lim_{|i-j|\to\infty}\mathcal{K}^y_{ij}$ in the large $|i-j|$ limit, describing the nontrivial chirality of the ground state. Through numerics, we also found that the
FLO phase is characterized by the algebraic decay of correlations distinguished from other magnetic phases in the
phase diagram Fig.~\ref{fig:phase}(a).

In the following, we will further discuss how to determine the phase boundary in the phase diagram Fig.~\ref{fig:phase}(a). We have calculated the entanglement entropy $S_{\rm E}=-{\rm tr}\rho_{\rm A}\ln \rho_{\rm A}$, where $\rho_{\rm A}$ is the reduced density matrix of a half chain. The transition point is determined with maximum $S_{\rm E}$
for each length, and then deduce the critical value in the thermodynamic limit by making an extrapolation with
respect to 1/L~\cite{2008_Amico_RevModPhys,2002_Osterloh_nature,2003_Vidal_PhysRevLett.90.227902,2004_Gu_PhysRevLett,
2002_Osborne_PhysRevA,2007_Chhajlany_PhysRevA}. Figure ~\ref{fig:entropy}(a) and (b) show such an extrapolation for
the critical orbital DM interaction strength with various chain lengths, which follows well a linear scaling behavior,
to determine the transition bewteen SFO and CHO phases. From numerics, we also confirm that the phase diagram obtained from the maximal entanglement entropy is consistent with that from analyzing the asymptotic behavior of correlation functions.

\textit{Tunability of the orbital exchange interaction $\raisebox{0.01mm}{---}$ } We now show how the relative
strength and sign of the different orbital exchange interactions in Eq.~\eqref{HMottSpinFinal}
can be controlled simply by tuning the external gradient magnetic field in our proposal. Here we restrict our discussion in the harmonic approximation. However, even under this simplistic assumption, our proposal still shows the great tunability as illustrated below. Under the harmonic approximation, the inter-orbital and intra-orbital bosonic interacting strength in our system satisfies the relation $U_{p_xp_x}=U_{p_yp_y}=3U_{p_xp_y}$. As shown in Fig. ~\ref{fig:coupling}, both the relative ratio and even the sign of the orbital exchange interactions in Eq.~\eqref{HMottSpinFinal} can be tuned via varying the hybridization strength between $p_x$ and $p_y$ orbitals characterized by the ratio ${t}/{t_y}$ here. To be more specific, when ${t}/{t_y}$ is small, the FM coupling for $S_{i}^{z}S_{j}^{z}$ in Eq.~\eqref{HMottSpinFinal} will dominate the system indicating
the FMO phase will be favored. While increasing ${t}/{t_y}$, the orbital DM interaction in Eq.~\eqref{HMottSpinFinal} will dominate the system and the CHO phase will be favored. Furthermore, the inset in Fig. ~\ref{fig:coupling} shows the ratio ${t}/{t_y}$ can be tuned through adjusting the external gradient magnetic field. This demonstrates that our proposal would pave an alternative way to realize and to further control the orbital magnetism, in particular, the chirality encoded within it.

\textit{Conclusion $\raisebox{0.01mm}{---}$} We have demonstrated a new approach to achieve the chiral orbital magnetism via the combination of inversion symmetry breaking and quantum degeneracy of orbital degrees of freedom in a cold atom based optical lattice system. This approach is rather generic to optical lattices than restricted to the setup considered in this work. Its principle is readily generalizable to higher dimensions with straightforward modifications, potentially circumventing the challenges in Raman-induced spin-orbit coupling scheme. This approach thus complements with a new window in cold gases to realize and furthermore to control the various orbital magnetism, in particular, the non-trivial chirality there.

\textit{Acknowledgment $\raisebox{0.01mm}{---}$} B. L. thanks helpful discussions with X. Li and W. V. Liu.
B. L. is supported by NSFC Grant No. 11774282. P. Z. is
supported by NSFC Grant No. 11604255. H. G. is supported
by NSFC Grant No. 11774286. F. L. is supported
by NSFC Grant No. 11534008 and the National Key R\&D
Project (Grant No. 2016YFA0301404).

\bibliographystyle{apsrev}
\bibliography{tFF}


\onecolumngrid

\renewcommand{\thesection}{S-\arabic{section}}
\setcounter{section}{0}  
\renewcommand{\theequation}{S\arabic{equation}}
\setcounter{equation}{0}  
\renewcommand{\thefigure}{S\arabic{figure}}
\setcounter{figure}{0}  

\indent

\begin{center}\large
\textbf{Supplementary Material:\\ Chiral orbital magnetism of $p$-orbital bosons in optical lattices}
\end{center}

\section{Effective spin Hamiltonian}

Here we will provide a detailed description of the procedure to derive the effective spin Hamiltonian
Eq. (2) from the Hubbard model Eq. (1) when considering the strong coupling limit and the unit filling
of the lattice sites. In such regime, to obtain the orbital exchange interactions in Eq. (2) resulting
from the virtual hopping processes, we define the $\hat{P}$ and $\hat{Q}=1-\hat{P}$ operators that project
into the subspace of states with a unit occupation and into the perpendicular subspace, respectively.
The eigenvalue problem associated to the model Hamiltonian $\hat{H}$ in Eq. (1) can be written as
\begin{equation}
\hat{H}(\hat{P}+\hat{Q})|\Psi \rangle=E|\Psi \rangle
\label{DeHam}
\end{equation}
where $|\Psi \rangle$ is the eigenstate with eigenenergy $E$ of the Hamiltonian $\hat{H}$. We then further
act with $\hat{P}$ and $\hat{Q}$ from the left of both sides in Eq.~\eqref{DeHam}. Then, the effective
Hamiltonian which describes the one particle Mott phase can be expressed as
\begin{equation}
\hat{H}_{eff}=-\hat{P}\hat{H}_{t}\hat{Q}\frac{1}{\hat{Q}\hat{H}_{U}\hat{Q}-E}
\hat{Q}\hat{H}_{t}\hat{P}
\label{effHam}
\end{equation}
where $\hat{H}_{U}$ and $\hat{H}_{t}$ are the interaction and hopping part
of the Hamiltonian in Eq. (1). Let us now consider the $2$-site problem and
define a basis for the doubly occupied states which can be spanned by a set
$\{ |p_{x}p_{x}\rangle \equiv 2^{-1/2}\hat{b}%
_{p_x}^{\dagger }\hat{b}_{p_x}^{\dagger }|0\rangle,$  $|p_{x}p_{y}\rangle \equiv \hat{b}_{p_x}^{\dagger }\hat{b}_{p_y}^{\dagger }|0\rangle,$
$|p_{y}p_{y}\rangle \equiv 2^{-1/2}\hat{b}_{p_y}^{\dagger }\hat{b}_{p_y}^{\dagger }|0\rangle
\}$, where
$|0\rangle$ is the empty state. Then $H_Q \equiv \hat{Q}\hat{H}_{U}\hat{Q}$ can be
represented in a matrix form
\begin{equation}
H_Q =\begin{pmatrix}
   U_{p_xp_x} & 0 &U_{p_xp_y} \\
   0& 2U_{p_xp_y} &0 \\
   U_{p_xp_y} &0 &  U_{p_yp_y}
\end{pmatrix}
\end{equation}
On this basis we arrive at the effective
spin Hamiltonian in Eq. (2) by computing the relevant matrix elements of Eq.~\eqref{effHam}
up to the second order virtual hopping processes.

\section{Dzyaloshinskii-Moriya (DM) interaction resulting from the combination of inversion symmetry breaking and quantum degeneracy of orbital degrees of freedom}

\begin{figure}[t]
\begin{center}
\includegraphics[width=8cm]{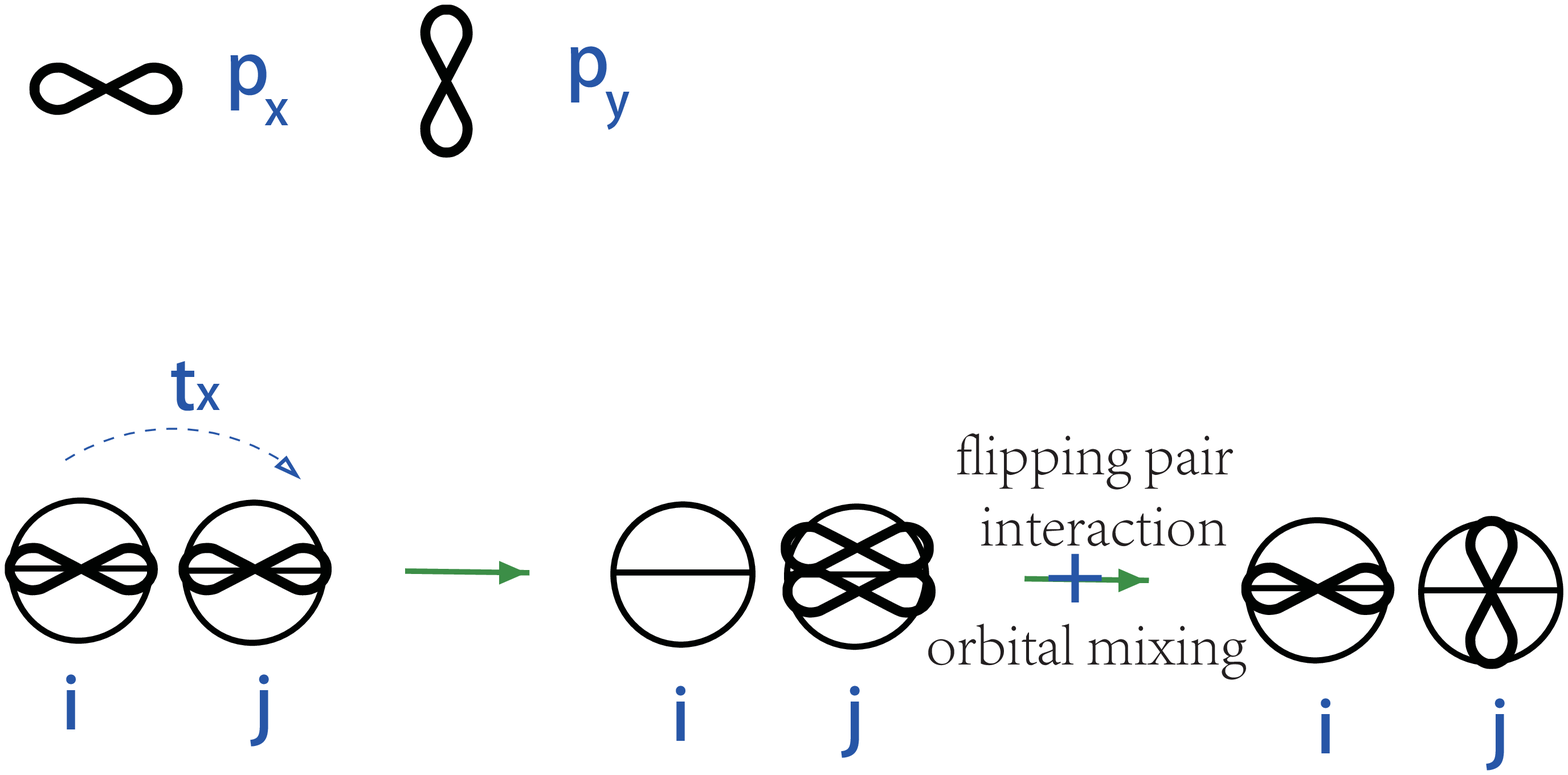}
\end{center}
\caption{Virtual hopping process giving rise to the orbital DM
interaction. $i$ and $j$ label two neighboring sites.} \label{fig:Vhopping}
\end{figure}

As illustrated in the previous section, the orbital exchange interactions can be understood from the virtual hopping processes, which can be described in terms of an effective Hamiltonian obtained from the perturbative expansion of the tunneling processes up to second order. Here we will describe how to understand the orbital Dzyaloshinskii-Moriya (DM) interaction resulting from the combination of inversion symmetry breaking and quantum degeneracy of orbital degrees of freedom. First of all, the inversion symmetry breaking along the $y$-direction will induce the non-trivial hybridization between the degenerate orbitals, i.e., $p_x$ and $p_y$ orbitals. Secondly, quantum degeneracy of orbital degrees of freedom
allows a special type interaction between spinless bosons (the last term in Eq. (1)) describing the flipping of a pair of bosonic atoms. Without either one of these two ingredients, the second order virtual hopping process shown in Figure ~\ref{fig:Vhopping} will not be allowed. However, it is worth to note that such a virtual hopping process will lead to the orbital Dzyaloshinskii-Moriya (DM) interaction. For example, for the process shown in Figure ~\ref{fig:Vhopping}, the effective Hamiltonian calculated from Eq.~\eqref{effHam} acquires a term of the form $-\sum_{<i,j>}\sum_{\alpha}\frac{2t_xtU_{p_xp_y}}{U^2}n_{\alpha}(\mathbf{x}_i)b^\dag_{\alpha}(\mathbf{x}_j)
b_{{\alpha}^{\prime }}(\mathbf{x}_j)$, which contributes the orbital DM interaction term in Eq. (2).

\end{document}